\documentclass[12pt]{article}
\usepackage{epsfig,amssymb,amsmath,psfrag,axodraw4j}

\def\Li{{\rm Li}}
\def\cN{{\mathcal N}}
\def\cO{{\mathcal O}}

\newcommand{\br}[1]{\langle #1 \rangle}
\newcommand{\homcord}[4]{\left[\begin{array}{c}#1\\#2\\#3\\#4\end{array}\right]}
\newcommand{\cS}{{\cal S}}

\def\beq{\begin{equation}}
\def\eeq{\end{equation}}

\def\bsp#1\esp{\begin{split}#1\end{split}}

\begin{document}

\catcode`\@=11
\font\manfnt=manfnt
\def\Watchout{\@ifnextchar [{\W@tchout}{\W@tchout[1]}}
\def\W@tchout[#1]{{\manfnt\@tempcnta#1\relax%
  \@whilenum\@tempcnta>\z@\do{%
    \char"7F\hskip 0.3em\advance\@tempcnta\m@ne}}}
\let\foo\W@tchout
\def\dubious{\@ifnextchar[{\@dubious}{\@dubious[1]}}
\let\enddubious\endlist
\def\@dubious[#1]{%
  \setbox\@tempboxa\hbox{\@W@tchout#1}
  \@tempdima\wd\@tempboxa
  \list{}{\leftmargin\@tempdima}\item[\hbox to 0pt{\hss\@W@tchout#1}]}
\def\@W@tchout#1{\W@tchout[#1]}
\catcode`\@=12


\thispagestyle{empty}

\begin{flushright}
HU-EP-11/22 \hskip1cm
CERN--PH--TH/2011-105\hskip1cm
SLAC--PUB--14458 \hskip1cm 
LAPTH-016/11 \\
IPPP/11/21\hskip1cm
DCPT/11/42\hskip1cm
NSF-KITP-11-072
\end{flushright}

\begingroup\centering
{\Large\bfseries\mathversion{bold}
The one-loop six-dimensional hexagon integral\\
with three massive corners \par}%
\vspace{8mm}

\begingroup\scshape\large
Vittorio Del Duca$^{(1,2)}$,
Lance~J.~Dixon$^{(3,4)}$, James~M.~Drummond$^{(4,5)}$\\
\large Claude Duhr$^{(6,2)}$, Johannes M.~Henn$^{(7,2)}$, Vladimir A. Smirnov$^{(8)}$\\
\endgroup
\vspace{6mm}
\begingroup\small
$^{(1)}$ \emph{INFN, Laboratori Nazionali Frascati, 00044 Frascati (Roma), Italy}\\
$^{(2)}$ \emph{Kavli Institute for Theoretical Physics, University of California, Santa Barbara, CA 93106, USA}\\
$^{(3)}$ \emph{SLAC National Accelerator Laboratory, Stanford University, Stanford, CA 94309, USA} \\
$^{(4)}$ \emph{PH-TH Division, CERN, Geneva, Switzerland} \\
$^{(5)}$ \emph{LAPTH, Universit\'e de Savoie, CNRS, B.P. 110, F-74941 Annecy-le-Vieux Cedex, France}\\
$^{(6)}$ \emph{Institute for Particle Physics Phenomenology, University of Durham, Durham, DH1 3LE, U.K.}\\
$^{(7)}$ \emph{Institut f\"ur Physik, Humboldt-Universit\"at zu Berlin, Newtonstra{\ss}e 15, D-12489 Berlin, Germany}\\
$^{(8)}$ \emph{Nuclear Physics Institute of Moscow State University, Moscow 119992, Russia}
\endgroup

\vspace{0.6cm}
\begingroup\small
E-mails:\\
{\tt delduca@lnf.infn.it}, {\tt lance@slac.stanford.edu}, {\tt drummond@lapp.in2p3.fr},\\
{\tt claude.duhr@durham.ac.uk}, {\tt henn@physik.hu-berlin.de}, {\tt smirnov@theory.sinp.msu.ru}
\endgroup
\vspace{1.2cm}

\textbf{Abstract}\vspace{5mm}\par
\begin{minipage}{14.7cm}
We compute the six-dimensional hexagon integral with three non-adjacent external masses
analytically. After a simple rescaling, it is given by a function of six dual conformally invariant 
cross-ratios. The result can be expressed as a sum of 24 terms involving only
one basic function, which is a simple linear combination of logarithms, dilogarithms, and 
trilogarithms of uniform degree three transcendentality.
Our method uses differential equations to determine the symbol of the function,
and an algorithm to reconstruct the latter from its symbol.
It is known that six-dimensional hexagon integrals are closely related
to scattering amplitudes in $\cN=4$ super Yang-Mills theory, and we therefore expect
our result to be helpful for understanding the structure of scattering amplitudes in 
this theory, in particular at two loops.
\end{minipage}\par
\endgroup

\newpage

\section{Introduction}

Scalar $n-$point integrals in dimensions $D>4$ are interesting objects for a number of reasons. 
They appear in the $\cO(\epsilon)$ part of $(D=4-2\epsilon)$-dimensional one-loop amplitudes \cite{Bern:1996ja},
which are required for computations at higher loop orders.

Quite generally, higher-dimensional scalar integrals are related to tensor integrals in 
$D=4$ dimensions \cite{Bern:1993kr}. In particular, the $D=6$ dimensional hexagons
are related to finite tensor integrals \cite{ArkaniHamed:2010gh} that appear in $\cN=4$ super Yang-Mills (SYM). 
More precisely, they 
appear as derivatives of four-dimensional two-loop tensor integrals. Moreover, applying
a further differential operator, the integrals reduce to four-dimensional one-loop tensor integrals 
\cite{Dixon:2011ng}. See Ref.~\cite{Drummond:2010cz} for related work on differential equations 
relevant for integrals in $\cN=4$ SYM.
 
Finite dual conformal invariant
functions~\cite{Broadhurst:1993ib,Drummond:2006rz} are also prototypes of functions
that can appear in the remainder function of MHV amplitudes and the ratio function of non-MHV amplitudes
in $\cN=4$ SYM \cite{Bern:2008ap,Drummond:2008aq,Drummond:2008vq}.
Recently, the massless and one-mass hexagon integrals in $D=6$ dimensions were computed in Refs.~\cite{Dixon:2011ng,DelDuca:2011ne,DelDuca:2011jm}.
It was noted that the massless hexagon integral in $D=6$ resembles very closely the analytical result of the 
two-loop remainder function for $n=6$ points \cite{DelDuca:2009au,DelDuca:2010zg,Goncharov:2010jf}.
In this note, we extend the computations of hexagon integrals in $D=6$ dimensions to the case
of three non-adjacent external masses.

Our strategy is the following. We derive simple differential equations that relate the three-mass hexagon to
known pentagon integrals. These differential equations, together with a boundary condition, completely determine
the answer in principle. We find it convenient to first compute the symbol \cite{symbols} of the answer, and then 
reconstruct the function from that symbol.  

\section{Integral representation and differential equations}
We consider the hexagon integral with three massive corners, 
\beq\label{defH9}
H_{9} := \int \frac{d^{6}x_{i}}{i \pi^3} \frac{1}{ x_{1i}^2 x_{2i}^2 x_{4i}^2 x_{5i}^2 x_{7i}^2 x_{8i}^2} \,,
\eeq
where we used dual (or region) coordinates
$p^{\mu}_{j} = x^{\mu}_{j} - x^{\mu}_{j+1}$ (with indices being defined modulo $9$), 
and $x^{\mu}_{ij} = x^{\mu}_{i} - x^{\mu}_{j}$.
The on-shell conditions read $x_{12}^2 =0, x_{45}^2=0$ and $x_{78}^2=0$.
As a scalar integral, $H_{9}$ is a function of the (non-zero) external Lorentz invariants $x_{jk}^2$.
We work in signature $({-}{+}{+}{+})$, so that
the Euclidean region has all (non-zero) $x_{jk}^2$ positive.

Dual conformal covariance~\cite{Broadhurst:1993ib,Drummond:2006rz} of $H_{9}$,
in particular under the inversion of all dual coordinates,
$x^{\mu} \to x^{\mu}/x^2$, allows us to write
\beq
H_{9} =: \frac{1}{x_{15}^2 x_{27}^2 x_{48}^2 } {\Phi}_{9}(u_{1},\ldots,u_6)\,,
\eeq
where the cross ratios
\beq\bsp
&u_1 := \frac{ x_{25}^2 x_{17}^2 }{x_{15}^2 x_{27}^2 }\,,\quad u_2 := \frac{ x_{58}^2 x_{41}^2 }{x_{48}^2 x_{15}^2 } \,,\quad u_3 := \frac{x_{82}^2 x_{74}^2}{x_{27}^2 x_{48}^2}\,,\\
&u_4 := \frac{ x_{24}^2 x_{15}^2 }{x_{14}^2 x_{25}^2 } \,,\quad u_5 := \frac{ x_{57}^2 x_{48}^2 }{x_{47}^2 x_{58}^2} \,,\quad u_6 := \frac{ x_{81}^2 x_{72}^2 }{x_{82}^2 x_{17}^2}\,,
\esp\eeq
are invariant under dual conformal transformations.
Furthermore, the one-loop hexagon integral with three non-adjacent masses is invariant under the action of the dihedral symmetry group $D_3\simeq S_3$, generated by the cyclic rotation $c$ and the reflection $r$ acting on the dual coordinates via
\beq\label{eq:dihed_sym}
x_{j}^\mu \stackrel{c}{\longrightarrow} x_{j+3}^\mu {\rm~~and~~} x_j^\mu \stackrel{r}{\longrightarrow} x_{9-j}^\mu\,,
\eeq
where as usual all indices are understood modulo 9. It is easy to see that under the symmetry the six conformal cross ratios group into two orbits of three elements,
\beq\begin{array}{cc}
\displaystyle u_1 \stackrel{c}{\longrightarrow} u_2 \stackrel{c}{\longrightarrow} u_3 \stackrel{c}{\longrightarrow}u_1 \,,&
\displaystyle u_4 \stackrel{c}{\longrightarrow} u_5 \stackrel{c}{\longrightarrow} u_6 \stackrel{c}{\longrightarrow}u_4 \,,\\
\displaystyle u_1 \stackrel{r}{\longleftrightarrow} u_3\,, & \displaystyle u_4 \stackrel{r}{\longleftrightarrow} u_5\,,\\
\displaystyle u_2 \stackrel{r}{\longleftrightarrow} u_2\,, &\displaystyle u_6 \stackrel{r}{\longleftrightarrow} u_6\,.
\end{array}\eeq

One can easily derive a differential equation for $H_9$ by noting that
\beq
(x_{21} \cdot \partial_{x_{2}} + 1) \frac{1}{x_{1i}^2 x_{2i}^2 } = \frac{1}{(x_{2i}^2)^2} \,.
\eeq
Applying this differential operator to Eq.~(\ref{defH9}), we find
\beq \label{diffeqH9}
(x_{21} \cdot \partial_{x_{2}} + 1) H_{9} = \int \frac{d^{6}x_{i}}{i \pi^3} \frac{1}{  (x_{2i}^2)^2 x_{4i}^2 x_{5i}^2 x_{7i}^2 x_{8i}^2}  =: P_{8} \,.
\eeq
The one-loop pentagon integral $P_{8}$ appearing as an inhomogeneous term in this equation is equivalent to a known four-dimensional pentagon integral \cite{Dixon:2011ng}\footnote{In Refs.~\cite{Dixon:2011ng,Drummond:2010cz}, the notation $\tilde{\Psi}$ was used for $\Psi_{8}$.},
\beq
P_{8}  =:  \frac{1}{x_{25}^2 x_{27}^2 x_{48}^2} {\Psi}_{8}(u_3 ,u_4 u_2 , u_5 )\,.
\eeq
The latter is given by
\beq\bsp
{\Psi}_{8}(u,v,w) \,=&\,\, \frac{1}{1-u-v+u v w} \big[ \log u \log v + \Li_{2}(1-u) + \Li_{2}(1-v) + \Li_{2}(1-w) \\
& \hskip3.1cm \, - \Li_{2}(1-u w) - \Li_{2}(1- vw) \big] \,.
\esp\eeq
We can rewrite Eq.~(\ref{diffeqH9}) as a differential equation for 
the rescaled hexagon integral ${\Phi}_{9}(u_{1},\ldots,u_6)$ that 
depends on cross-ratios only,
\beq \label{diffeqphi9}
D_1 {\Phi}_{9}(u_1,\ldots,u_6) ={\Psi}_{8}(u_3 ,u_4 u_2 , u_5 )\,,
\eeq
where
\beq
D_1 := u_1 +u_1 u_6 (u_6 -1) \partial_6 + (u_4 -1)\partial_4 + u_1 (u_1 -1) \partial_1 + u_1 (1-u_6 )u_3 \partial_3 \,,
\eeq
with $\partial_i := \partial/\partial u_i$.
By cyclic and reflection symmetry, 
we have a total of six differential equations. 
It turns out that only five of them are independent.
The remaining freedom can be fixed, \emph{e.g.}, by the boundary condition $H_{9}(u_1,u_2,u_3,0,0,0)=H_{6}(u_1,u_2,u_3)$,
with $H_{6}$ given explicitly in Refs.~\cite{Dixon:2011ng,DelDuca:2011ne}. (Alternatively, one could derive further differential equations,
as in Ref.~\cite{Dixon:2011ng}).
Therefore, the set of equations and the boundary condition completely determine $H_9 $. 

In the next section, we will use this set of differential equations to determine the symbol ${\cal S}(\tilde{\Phi}_{9})$,
where $\tilde{\Phi}_{9}$ is obtained from ${\Phi}_{9}$ by a simple rescaling, see Eq.~(\ref{phi-rescaling}).
Then, we will reconstruct the function $\tilde{\Phi}_{9}$ (and equivalently $H_{9}$) from its symbol.

\begin{figure}
\psfrag{x1}[cc][cc]{$\scriptstyle x_1$}
\psfrag{x2}[cc][cc]{$\scriptstyle x_2$}
\psfrag{x7}[cc][cc]{$\scriptstyle x_7$}
\psfrag{x4}[cc][cc]{$\scriptstyle x_4$}
\psfrag{x5}[cc][cc]{$\scriptstyle x_5$}
\psfrag{x8}[cc][cc]{$\scriptstyle x_8$}
\psfrag{xi1}[cc][cc]{$\scriptstyle y_1$}
\psfrag{xi3}[cc][cc]{$\scriptstyle y_4$}
\psfrag{xi5}[cc][cc]{$\scriptstyle y_7$}
\psfrag{equal}[cc][cc]{$=$}
\psfrag{two}[cc][cc]{$\scriptstyle 2$}
\psfrag{one}[cc][cc]{$\scriptstyle 1$}
\psfrag{a}[cc][cc]{(a)}
\psfrag{b}[cc][cc]{(b)}
\centerline{
 {\epsfxsize12cm  \epsfbox{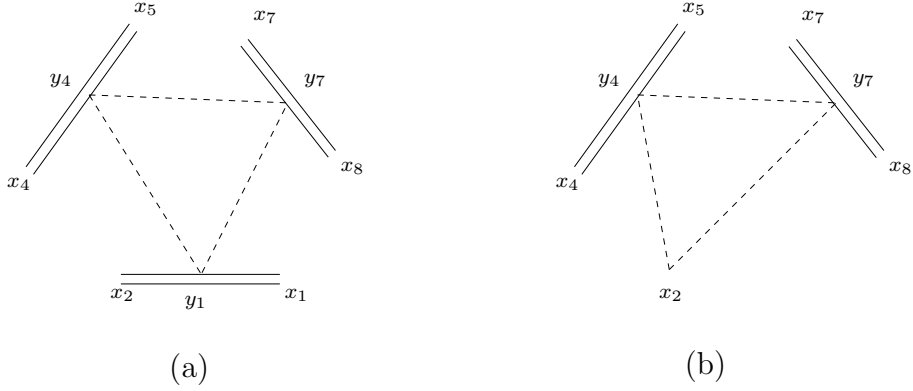}}
}
\caption{\small
(a) depicts the representation of $H_{9}$ as a line integral, see Eqs.~(\ref{H9line}) and (\ref{H9line2}).
The differential operator in Eq.~(\ref{diffeqH9}) localizes the $y_{1}$ integration to $x_{2}$, yielding $P_{8}(x_{2},x_{4},x_{5},x_{7},x_{8})$, see (b).}
\label{fig-wl-diff}
\end{figure}

We note that there is a simple line integral representation of $H_{9}$ \cite{Dixon:2011ng}, see Fig.~\ref{fig-wl-diff}(a),
\beq\label{H9line}
H_{9} = \int_{0}^{1} d\xi_{1} d\xi_{4} d\xi_{7} \frac{1}{(y_{1}-y_{4})^2 (y_{4}-y_{7})^2 (y_{7}-y_{1})^2 }\,,
\eeq
where $y_{1}^{\mu} = x_{1}^{\mu} + \xi_{1} x_{21}^{\mu}$,  $y_{4}^{\mu} = x_{4}^{\mu} + \xi_{4} x_{54}^{\mu}$ and $y_{7}^{\mu} = x_{7}^{\mu} + \xi_{7} x_{87}^{\mu}$.
The pentagon integral $P_{8}$ can be expressed in a similar way, which allows us to write
\beq\label{H9line2}
H_{9} = \int_{0}^{1} d\xi_{1} P_{8}( y_{1}(\xi_{1}), x_{4},x_{5},x_{7},x_{8} )\,.
\eeq
In this form, the differential equation (\ref{diffeqH9}) has the interpretation of localizing one of the line integrals,
in this case $y_{1}(\xi_{1}) \to x_{2}$, see Fig.~\ref{fig-wl-diff}(b). 
It is interesting that similar integrals where certain propagators are localized at cusp
points have also appeared in computations of two-loop Wilson loops \cite{Drummond:2007au}.

From this discussion it is also clear that the integral reduces further in degree under the action of other differential operators, 
until one eventually obtains a rational function. More explicitly, the operator
$(x_{54} \cdot \partial_{x_5} + 1)$ acting on $P_8 $ similarly gives a first-order differential equation relating ${\Psi}_{8}$ 
to a single-log function, namely a $3$-mass box integral with two doubled propagators,
\beq
X_{7} := \int \frac{d^{6}x_{i}}{i \pi^3} \frac{1}{  (x_{2i}^2)^2 (x_{4i}^2 )^2 x_{7i}^2 x_{8i}^2}  =:
\frac{1}{x_{25}^2 x_{27}^2 x_{58}^2} \chi_7(u_3 u_5)
\eeq
where $\chi_{7}(y) = \log(y)/(y-1)$.
Acting further on $X_7 $ with $(x_{87} \cdot \partial_{x_8} + 1)$ gives the 3-mass
triangle with three doubled propagators, which is a constant up to the usual
prefactors, $1/(x_{25}^2 x_{58}^2 x_{82}^2)$.

The representation (\ref{H9line}) may also be useful for numerical checks. For future reference,
it can be rewritten as
\beq\bsp\label{H9param}\Phi_{9}&(u_1,\ldots,u_6) \\
&\,= \int_{0}^{1} \frac{ d\xi_{1} d\xi_{4} d\xi_{7}}{(u_2 \bar{\xi}_{1} \bar{\xi}_{4} + u_4 u_2 \xi_1 \bar{\xi}_4 + \xi_4 )(u_3 \bar{\xi}_{4} \bar{\xi}_{7} + u_5 u_3 \xi_4 \bar{\xi}_7 + \xi_7 ) 
(u_1 \bar{\xi}_{7} \bar{\xi}_{1} + u_6 u_1 \xi_7 \bar{\xi}_1 + \xi_1 )} \,,
\esp\eeq
where $\bar{\xi}_i := 1 - \xi_i$.

\section{Symbols from differential equations}
\label{sec:symb_diff_eq}

We find that the following definition 
\beq \label{phi-rescaling}
{\Phi}_{9}(u_{1},\ldots,u_{6}) =:  \frac{1}{\sqrt{\Delta_{9}}} \tilde{\Phi}_{9}(u_{1},\ldots,u_{6}) \,.
\eeq
leads to a pure function $\tilde{\Phi}_{9}(u_{i})$, \emph{i.e.}, a function that can be written as a linear
combination of transcendental functions, with numerical coefficients only.
Here
\beq\bsp
\Delta_9 :=&\, (1-u_1 - u_2 - u_3 + u_4 u_1 u_2 + u_5 u_2 u_3 + u_6 u_3 u_1 - u_1 u_2 u_3 u_4 u_5 u_6 )^2 \\
&\,- 4 u_1 u_2 u_3 (1-u_4 )(1-u_5 ) (1-u_6 )\,.
\esp\eeq
Using this definition, and $D_1(1/\sqrt{\Delta_9})=0$, we can rewrite Eq.~(\ref{diffeqphi9}) as
\beq\label{diffeqphi9tilde}
\tilde{D}_{1}   \tilde{\Phi}_{9}(u_{1},\ldots,u_{6}) = \tilde{\Psi}_{8}(u_3 ,u_4 u_2 , u_5 ) \,,
\eeq
where 
\beq\bsp
\tilde{D}_{1} :=&\, \frac{1}{\sqrt{\Delta}_{9}} (1-u_3 - u_2 u_4 + u_2 u_3 u_4 u_5 )  \, \times  \\
&\, \times \left[ u_1 u_6 (u_6 -1) \partial_6 + (u_4 -1)\partial_4 + u_1 (u_1 -1) \partial_1 + u_1 (1-u_6 )u_3 \partial_3 \right] \\
 =&\, \frac{1}{\sqrt{\Delta}_{9}} (1-u_3 - u_2 u_4 + u_2 u_3 u_4 u_5 ) (D_1 - u_1) \,,
\esp\eeq
and
\beq
\tilde{\Psi}_{8}(u,v,w):=(1-u-v+u v w)\,{\Psi}_{8}(u,v,w)\,.
\eeq
We find it convenient to convert (\ref{diffeqphi9tilde}) into a differential equation for the symbol of $\tilde{\Phi}_{9}$,
which reads
\beq\label{diffeqphi9tildesymbol}
\tilde{D}_{1} {\cal S}(\tilde{\Phi}_{9})(u_{1},\ldots,u_{6}) ={\cal S}(\tilde{\Psi}_{8})(u_3 ,u_4 u_2 , u_5 ) \,.
\eeq
Here the differentiation of a symbol is defined by
\beq
\partial_{x} \, \left( a_{1} \otimes \ldots 
             \otimes a_{n-1} \otimes a_{n} \right)
 = \partial_{x} \log(a_{n}) \, \times \, a_{1} \otimes \ldots
             \otimes a_{n-1} \,.
\eeq
The following set of variables is useful to describe the solution,
\beq
W_i  := \frac{g_i - \sqrt{\Delta_{9}}}{g_i  + \sqrt{\Delta_{9}}}  \,,\quad i=1\ldots 6\,,
\eeq
where
\beq\bsp
g_1 &\,:= 1 - u_1 - u_2 + u_3 + u_1 u_2 u_4 - u_2 u_3 u_5 - 2 u_3 u_6 + u_1 u_3 u_6 + 2 u_2 u_3 u_5 u_6 - u_1 u_2 u_3 u_4 u_5 u_6 \,, \\
g_4 &\,:= 1 - u_1 - u_2  - u_3 + 2 u_1 u_2 - u_1 u_2 u_4 + u_2 u_3 u_5 + u_1 u_3 u_6 - 
 2 u_1 u_2 u_3 u_5 u_6 + u_1 u_2 u_3 u_4 u_5 u_6  \,,\nonumber
\esp\eeq
and where $g_2 , g_3$ ($g_5 , g_6 $) are  obtained from $g_1 $ ($g_4 $) by cyclic mappings $1\to 2 \to 3 \to1; 4 \to 5 \to 6 \to 4$.
These variables have a nice behavior under the differential operators, \emph{e.g.},
\beq
\tilde{D}_1\log(W_i) = \left\{\begin{array}{ll}
-1, & \textrm{if } i=6\\
0, & \textrm{otherwise}\end{array}\right.
{\rm~~and~~}
\tilde{D}_4\log(W_i) = \left\{\begin{array}{ll}
1, & \textrm{if } i=1\\
-1, & \textrm{if } i=2\textrm{ or } 4\\
0, & \textrm{otherwise}\end{array}\right.\,,
\eeq
where $\tilde{D}_4$ is defined as the image of $\tilde{D}_1$ under the reflection $u_4\leftrightarrow u_6$ and  $u_2\leftrightarrow u_3$.
Given these variables, we can write the solution to Eq.~(\ref{diffeqphi9tildesymbol}) as
\beq
{\cal S}(\tilde{\Phi}_{9})(u_{1},\ldots,u_{6}) = {\cal S}(\tilde{\Psi}_{8})(u_3 ,u_4 u_2 , u_5 )  \otimes W_6 + T\,,
\eeq
where $T$ satisfies $\tilde{D}_{1} T =0$.
Taking into account the differential equations related to (\ref{diffeqphi9tildesymbol}) by symmetry further
restricts the form of $T$. The particular solution we obtained is in general not an integrable symbol. We therefore proceed and add a particular $T_h $ satisfying 
$\tilde{D}_{i} T_h  =0$ (for $i=1\ldots 5$) to obtain an integrable symbol. 
Finally, additional terms satisfying the homogeneous equations $\tilde{D}_{i} T =0$ are fixed by
demanding that the symbol for $\tilde{\Phi}_{6}$ for the massless hexagon~\cite{Dixon:2011ng,DelDuca:2011ne} is reproduced when $u_4 = u_5 = u_6 =0$.

Following this procedure, we find that the symbol ${\cal S}(\tilde{\Phi}_{9})$ can then be written as
\beq\label{eq:symbol_Phi9}
{\cal S}(\tilde{\Phi}_{9})=  \sum_{i=1}^{6}  {\mathcal{S}}(f_{i}) \otimes W_{i}\,,
\eeq
where $f_{i}$ are the following degree three functions,
\beq\bsp\label{eq:f_functions}
f_1 &\,:= \tilde{\Psi}_{8}(u_2, u_1 u_6, u_4 ) + \tilde{\Psi}_{8}(u_1, u_2 u_5, u_4 ) +  \tilde{\Psi}_{8}(u_2, u_3 u_6, u_5 ) - F(u_1 , u_2 ,u_3 ,u_4 , u_5 , u_6 )\,,\\
f_4 &\,:= -\tilde{\Psi}_{8}(u_1 ,u_3 u_5 ,u_6) \,.
\esp\eeq
Here the quantities $f_2$, $f_3$ ($f_5 , f_6 $) are  obtained from $f_1 $ ($f_4 $) by cyclic mappings $1\to 2 \to 3 \to1; 4 \to 5 \to 6 \to 4$.
Moreover,
\beq\bsp
F :=
 &\, 2  \tilde{\Psi}_{8}(u_1 ,u_2 ,u_4)  + \log u_1 \log u_5 + \log u_2 \log u_6 - \log u_3 \log u_4\,.
\esp\eeq
Note that one can rearrange terms in Eq.~\eqref{eq:f_functions} because of the identity,
\beq\bsp
0 =&\,\tilde{\Psi}_{8}(u_3 ,u_2 u_4 ,u_5) +\tilde{\Psi}_{8}(u_1 ,u_3 u_5 ,u_6) + \tilde{\Psi}_{8}(u_2 ,u_1 u_6 ,u_4)  \\
&\, - \tilde{\Psi}_{8}(u_3 ,u_1 u_4 ,u_6)   - \tilde{\Psi}_{8}(u_1 ,u_2 u_5 ,u_4) - \tilde{\Psi}_{8}(u_2 ,u_3 u_6 ,u_5) \,.
\esp\eeq

\section{Twistor geometry associated to a three-mass hexagon}
\label{sec:twistor}
The differential equation technique allowed us to obtain the symbol of the one-loop three-mass hexagon integral. If we want to find the analytic expression for the integral, we need to integrate the symbol to a function.
We follow here the approach of Ref.~\cite{Gangl:2011}, which, after making a suitable choice for the functions that should appear in the answer, allows us to reduce the problem of integrating the symbol to a problem of linear algebra. The algorithm of Ref.~\cite{Gangl:2011}, however, requires the arguments of the symbol to be rational functions (of some parameters). From Eq.~\eqref{eq:symbol_Phi9} it is clear that in our case this requirement is not immediately fulfilled, because the variables $W_i$ are algebraic functions of the cross ratios $u_i$. In order to bypass this problem, we have to parametrize the six cross ratios such that $\Delta_9$ becomes a perfect square.

A convenient way to find a parametrization that turns $\Delta_9$ into a perfect square is to write the six cross ratios as ratios of twistor brackets. Indeed, even though we work in $D=6$ dimensions where the link to twistor space is not immediately obvious, we can nevertheless consider the cross ratios as being parametrized by cross ratios in twistor space $\mathbb{CP}^3$, because the functional dependence of $\Phi_9$ is only through the six conformally invariant quantities $u_i$, which do not make reference to the six-dimensional space. In other words, we can consider the external momenta to lie in a four-dimensional subspace, even as we integrate over six components of loop momentum.  Furthermore, in Ref.~\cite{Goncharov:2010jf} it was noted that in terms of momentum twistor variables, the equivalent of $\Delta_9$ in the massless case becomes a perfect square. Hence, momentum twistors seem to provide a natural framework to search for a suitable parametrization. We therefore briefly review the geometry of a three-mass hexagon configuration in momentum twistor space.

\begin{figure}[!t]
\begin{center}
\begin{picture}(379,125) (47,-67)
    \SetWidth{0.5}
    \SetColor{Black}
    \Text(63,19)[lb]{\normalsize{\Black{$x_1$}}}
    \Text(116,37)[lb]{\normalsize{\Black{$x_2$}}}
    \Text(168,19)[lb]{\normalsize{\Black{$x_4$}}}
    \Text(168,-37)[lb]{\normalsize{\Black{$x_5$}}}
    \Text(111,-65)[lb]{\normalsize{\Black{$x_7$}}}
    \Text(63,-37)[lb]{\normalsize{\Black{$x_8$}}}
    \SetWidth{2.0}
    \Line(90,28)(143,28)
    \Line(69,-10)(90,27)
    \Line(164,-10)(143,27)
    \Line(90,-46)(69,-9)
    \Line(90,-46)(143,-46)
    \Line(143,-46)(164,-9)
    \Line(90,28)(79,47)
    \Line(149,-65)(143,-46)
    \Line(185,-10)(164,-10)
    \Line(69,-10)(49,-14)
    \Line(155,-61)(143,-46)
    \Line(69,-10)(50,-4)
    \Line(90,-46)(79,-65)
    \Line(148,47)(142,28)
    \Line(156,43)(143,28)
    \Text(249,-28)[lb]{\normalsize{\Black{$Z_8$}}}
    \SetWidth{1.0}
    \Vertex(288,24){2.236}
    \Text(246,13)[lb]{\normalsize{\Black{$Z_9$}}}
    \Text(283,29)[lb]{\normalsize{\Black{$Z_1$}}}
    \Text(330,29)[lb]{\normalsize{\Black{$Z_2$}}}
    \SetWidth{2.0}
    \Line(246,3)(288,24)
    \Line[dash,dashsize=2](309,34)(330,45)
    \Line[dash,dashsize=2](225,-8)(246,3)
    \Line(288,24)(309,35)
    \SetWidth{1.0}
    \Vertex(253,6){2.236}
    \SetWidth{2.0}
    \Line(277,24)(340,24)
    \Line[dash,dashsize=2](341,24)(362,24)
    \Line[dash,dashsize=2](256,24)(277,24)
    \SetWidth{1.0}
    \Vertex(331,24){2.236}
    \Text(354,-32)[lb]{\normalsize{\Black{$Z_5$}}}
    \Text(391,-15)[lb]{\normalsize{\Black{$Z_4$}}}
    \Text(388,24)[lb]{\normalsize{\Black{$Z_3$}}}
    \SetWidth{2.0}
    \Line(383,-19)(383,34)
    \Line(352,-19)(394,2)
    \Line[dash,dashsize=2](331,-29)(352,-18)
    \Line[dash,dashsize=2](394,3)(415,14)
    \Line[dash,dashsize=2](383,-19)(383,-40)
    \Line[dash,dashsize=2](383,55)(383,34)
    \SetWidth{1.0}
    \Vertex(383,-2){2.236}
    \Vertex(352,-19){2.236}
    \Vertex(383,24){2.236}
    \Vertex(267,-19){2.236}
    \SetWidth{2.0}
    \Line(277,-40)(288,-40)
    \Text(277,-57)[lb]{\normalsize{\Black{$Z_7$}}}
    \Text(320,-58)[lb]{\normalsize{\Black{$Z_6$}}}
    \Line(298,-50)(266,-18)
    \Line[dash,dashsize=2](256,-8)(267,-19)
    \Line[dash,dashsize=2](298,-50)(309,-61)
    \Line(288,-40)(330,-40)
    \Line[dash,dashsize=2](331,-40)(352,-40)
    \Line[dash,dashsize=2](256,-40)(277,-40)
    \SetWidth{1.0}
    \Vertex(288,-40){2.236}
    \Vertex(320,-40){2.236}
  \end{picture}
\end{center}
\caption{\label{fig:twistor_space}The one-loop three-mass hexagon integral (left), and its geometric configuration in momentum twistor space $\mathbb{CP}^3$ (right). Only the intersection points $Z_1$, $Z_4$ and $Z_7$ have an intrinsic geometrical meaning, whereas all other twistors can be moved freely along the lines.}
\end{figure}
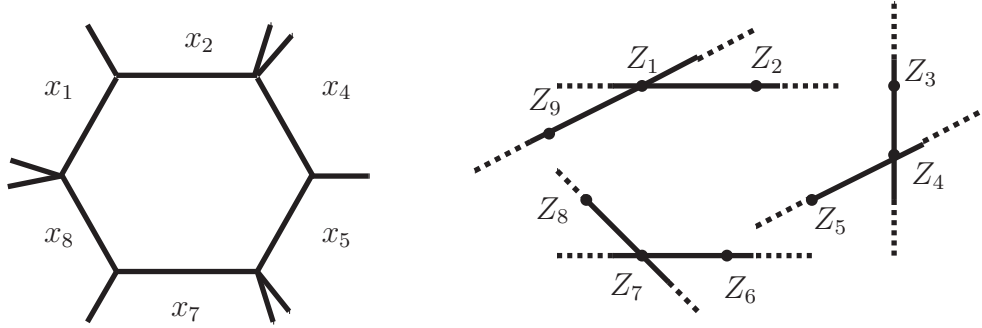

In order to describe this geometry, we assume that the dual coordinates $x_i$ are elements of four-dimensional Minkowski space $\mathbb{M}^4$. As the dependence of $\Phi_9$ is solely through cross ratios, we can assume that this condition is satisfied, as long as the `projection' to the four-dimensional space leaves the cross ratios invariant. The twistor correspondence then associates to each point $x_i$ in $\mathbb{M}^4$ a projective line $X_i$ in momentum twistor space, and two points $x_i$ and $x_j$ in $\mathbb{M}^4$ are lightlike separated if and only if the corresponding lines $X_i$ and $X_j$ intersect. In our case this implies that the six lines must intersect pairwise (See Fig.~\ref{fig:twistor_space}). Denoting the intersection points by $Z_1$, $Z_4$ and $Z_7$, we can define six more twistors by 
\beq\label{eq:XZZ}
X_i = Z_i\wedge Z_{i-1}, \quad i\in\{1,2,4,5,7,8\}\,.
\eeq
Note that the only points in twistor space that have an intrinsic geometric meaning are $Z_1$, $Z_4$ and $Z_7$, whereas the other six points are defined through Eq.~\eqref{eq:XZZ}, which is left invariant by the redefinitions
\beq\bsp\label{eq:shift_invariance}
& Z_2 \to Z_2 + \alpha_2 Z_1\,,\quad Z_5 \to Z_5 + \alpha_5 Z_4\,, \quad Z_8 \to Z_8 + \alpha_8 Z_7\,,\\
& Z_9 \to Z_9 + \alpha_9 Z_1\,,\quad Z_3 \to Z_3 + \alpha_3 Z_4\,, \quad Z_6 \to Z_6 + \alpha_6 Z_7\,,
\esp\eeq
where $\alpha_i$ are non-zero complex numbers.
These shifts simply express the fact that we can move the points along the line without altering the geometric configuration. Furthermore, the intersection of two lines $X_i$ and $X_j$ can be expressed through the condition,
\beq
\langle X_iX_j\rangle := \langle (i-1)\,i\,(j-1)\,j\rangle = \langle Z_{i-1}\,Z_i\,Z_{j-1}\,Z_j\rangle = \epsilon_{IJKL}\,Z_{i-1}^I\,Z_i^J\,Z_{j-1}^K\,Z_j^L = 0\,.
\eeq
Using the twistor brackets, the cross ratios $u_i$ can be parametrized as 
\beq\bsp
&u_1 = {\br{X_2X_5}\br{X_1X_7}\over\br{X_1X_5}\br{X_2X_7}}\,,\qquad
u_2 = {\br{X_5X_8}\br{X_4X_1}\over\br{X_4X_8}\br{X_1X_5}}\,,\qquad
u_3 = {\br{X_8X_2}\br{X_7X_4}\over\br{X_2X_7}\br{X_4X_8}}\,,\\
&u_4 = {\br{X_2X_4}\br{X_1X_5}\over\br{X_1X_4}\br{X_2X_5}}\,,\qquad
u_5 = {\br{X_5X_7}\br{X_4X_8}\over\br{X_4X_7}\br{X_5X_8}}\,,\qquad
u_6 = {\br{X_8X_1}\br{X_7X_2}\over\br{X_8X_2}\br{X_1X_7}}\,.
\esp\eeq
It is clear that the dihedral symmetry of the integral is reflected at the level of the twistors by
\beq\label{eq:twist_sym}
Z_i \stackrel{c}{\longrightarrow} Z_{i+3} {\rm~~and~~} Z_{i}\stackrel{r}{\longrightarrow} Z_{8-i}\,,
\eeq
where again all indices are understood modulo 9.
This action on the twistors induces an action on the lines $X_i$ and the planes $\overline{Z}_i = Z_{i-1}\wedge Z_i\wedge Z_{i+1}$ via
\beq\bsp
X_i \stackrel{c}{\longrightarrow} X_{i+3} &{\rm~~and~~} X_{i}\stackrel{r}{\longrightarrow} -X_{9-i}\,,\\
\overline{Z}_i \stackrel{c}{\longrightarrow} \overline{Z}_{i+3} &{\rm~~and~~} \overline{Z}_{i}\stackrel{r}{\longrightarrow} -\overline{Z}_{8-i}\,.
\esp\eeq

We now choose a particular representation for the twistors. Since the points $Z_1$, $Z_4$ and $Z_7$ play a special role, we choose their homogeneous coordinates as
\beq
Z_1=\homcord{0}{1}{0}{0}\,,\qquad Z_4=\homcord{0}{0}{1}{0}\,,\qquad Z_7=\homcord{0}{0}{0}{1}\,.
\eeq
As the other six points do not carry any intrinsic geometric meaning, we prefer not to fix them, but choose their homogeneous coordinates to be
\beq\label{eq:twist_param}
Z_i = \homcord{1}{x_i}{y_i}{z_i}\,,\qquad \textrm{ for } i\in \{2,3,5,6,8,9\}\,.
\eeq
(The $x_i$ and $y_i$ defined here should not be confused with the previous definitions, where they
were dual coordinates.)  In this parametrization the cross ratios then take the form
\beq\bsp\label{eq:cr_param}
& u_1 = \frac{\left(y_9-y_6\right) \left(z_2-z_5\right)}{\left(y_2-y_6\right) \left(z_9-z_5\right)}\,,\quad
u_2 = \frac{\left(x_5-x_8\right) \left(z_3-z_9\right)}{\left(x_3-x_8\right) \left(z_5-z_9\right)}\,,\quad
u_3 = \frac{\left(x_6-x_3\right) \left(y_8-y_2\right)}{\left(x_8-x_3\right) \left(y_6-y_2\right)}\,,\\
& u_4 = \frac{\left(z_2-z_3\right) \left(z_9-z_5\right)}{\left(z_2-z_5\right) \left(z_9-z_3\right)}\,,\quad
u_5 = \frac{\left(x_5-x_6\right) \left(x_3-x_8\right)}{\left(x_3-x_6\right) \left(x_5-x_8\right)}\,,\quad
u_6 = \frac{\left(y_6-y_2\right) \left(y_8-y_9\right)}{\left(y_8-y_2\right) \left(y_6-y_9\right)}\,.
\esp\eeq
Note that the cross ratios only depend on 12 out of the 18 homogeneous coordinates defined in Eq.~\eqref{eq:twist_param}, which is  a consequence of the shift invariance~\eqref{eq:shift_invariance}. The action of the dihedral symmetry that permutes the cross ratios is implemented in this parametrization via
\beq\bsp\label{eq:param_sym}
x_{i} \stackrel{c}{\longrightarrow} y_{i+3} \stackrel{c}{\longrightarrow} z_{i+6} \stackrel{c}{\longrightarrow} x_i {\rm~~~~~and~~~~~} x_{i} \stackrel{r}{\longleftrightarrow} z_{8-i} {\rm~~and~~} y_{i} \stackrel{r}{\longleftrightarrow} y_{8-i}\,.
\esp\eeq
This action seems to be inconsistent with Eq.~\eqref{eq:twist_sym}. However, we have broken the symmetry by freezing $Z_1$, $Z_4$ and $Z_7$ to constant values, and the symmetry is now reflected at the level of the cross ratios via Eq.~\eqref{eq:param_sym}. Finally, we note that $\Delta_9$ becomes a perfect square in these variables,
\beq\label{eq:Delta9_xyz}
\Delta_9 = \frac{\left(\left(x_6-x_8\right) \left(y_9-y_2\right) \left(z_3-z_5\right)+\left(x_5-x_3\right) \left(y_8-y_6\right) \left(z_2-z_9\right)\right)^2}{\left(x_3-x_8\right)^2 \left(y_6-y_2\right)^2 \left(z_9-z_5\right)^2}\,,
\eeq
and Eq.~\eqref{eq:Delta9_xyz} is manifestly invariant under the transformations~\eqref{eq:param_sym}. Having obtained a parametri\-zation that makes $\Delta_9$ into a perfect square, we can write the symbol in a form in which all the entries are rational functions of the variables we just defined, and hence the symbol now takes a form which allows it to be integrated using the algorithm of Ref.~\cite{Gangl:2011}. Furthermore, using this parametrization it is trivial to check that the symbol of $\tilde\Phi_9$ obtained in the previous section has the correct dihedral symmetry. In particular, we find that
\beq
c[\cS(\tilde\Phi_9)] = \cS(\tilde\Phi_9) {\rm~~and~~} r[\cS(\tilde\Phi_9)] = -\cS(\tilde\Phi_9)\,.
\eeq
The parametrization~\eqref{eq:cr_param} also makes it very easy to check the various soft limits of $H_9$. Indeed, we have
\beq\bsp
u_4 \to 0 \, \Leftrightarrow z_3 \to z_2\,,\qquad
u_5 \to 0 \, \Leftrightarrow x_6 \to x_5\,,\qquad u_6 \to 0 &\, \Leftrightarrow y_9 \to y_8\,.
\esp\eeq
We checked that in taking these limits $\cS(\tilde\Phi_{9})$ reduces to the symbols for the massless and one-mass hexagon integrals~\cite{Dixon:2011ng,DelDuca:2011ne,DelDuca:2011jm}.

\section{Integrating the symbol: the one-loop three-mass\\ hexagon integral}
\label{section:result}

As the parametrization of the cross ratios in terms of momentum twistors introduced in the previous section turns $\Delta_9$ into a perfect square, we can now integrate the symbol using the algorithm of Ref.~\cite{Gangl:2011}. However, even though the parametrization~\eqref{eq:cr_param} makes all the symmetries manifest, it uses a redundant set of parameters. We therefore choose a minimal set of parameters by breaking the $S_3$ symmetry down to its alternating subgroup $A_3\simeq \mathbb{Z}_3$ by fixing six of the twelve parameters,
\beq
x_6 = y_9 = z_3 = 0   {\rm~~and~~} x_3 = y_6 = z_9 = 1\,.
\eeq
The cross ratios then take the form
\beq\bsp\label{parametrization}
& u_1 = \frac{z_2-z_5}{\left(1-y_2\right) \left(1-z_5\right)}\,,\quad
 u_2 = \frac{x_5-x_8}{\left(1-x_8\right) \left(1-z_5\right)}\,,\quad
u_3 = \frac{y_8-y_2}{\left(1-x_8\right) \left(1-y_2\right)}\,,\\
& u _4 = \frac{z_2 \left(1-z_5\right)}{z_2-z_5}\,,\quad\quad\quad\,\,\,
u_5 = \frac{x_5 \left(1-x_8\right)}{x_5-x_8}\,,\quad\quad\quad\,\,\,
u_6 = \frac{y_8\left(1-y_2\right)}{y_8-y_2}\,,
\esp\eeq
and $\Delta_9$ can now be written as
\beq
\Delta_9 = \frac{\left(x_8 y_2 z_5 + (1-x_5) \left(1-y_8\right) \left(1-z_2\right)\right)^2}{\left(1-x_8\right)^2 \left(1-y_2\right)^2 \left(1-z_5\right)^2}\,.
\eeq
We note in passing that the Jacobian of the parametrization~\eqref{parametrization} is non-zero for generic values of the parameters.

In a nutshell, the algorithm of Ref.~\cite{Gangl:2011} proceeds in two steps:
\begin{enumerate}
\item Given the symbol of $\tilde\Phi_9$ computed in Section~\ref{sec:symb_diff_eq}, it constructs a set of rational functions $\{R_i(x_5,x_8,y_2,y_8,z_2,z_5)\}$ such that, \emph{e.g.}, symbols of the form $\cS(\textrm{Li}_n(R_i))$ span the vector space of which $\cS(\tilde\Phi_9)$ is an element.
\item Once a suitable set of rational functions has been obtained, it makes an ansatz
\beq\bsp
\tilde \varphi =&\, \sum_i c_i\,\textrm{Li}_3(R_i) + \sum_{i,j} c_{ij}\,\textrm{Li}_2(R_i)\,\log R_j + 
\sum_{i,j,k} c_{ijk}\,\log R_i\,\log R_j\,\log R_k\,,
\esp\eeq
where the $c_i$, $c_{ij}$ and $c_{ijk}$ are rational numbers to be determined such that 
\beq
\cS(\tilde\varphi) = \cS(\tilde\Phi_9)\,. 
\eeq
As the objects appearing in this last equation are tensors (\emph{i.e.}, elements of a vector space), the coefficients $c_i$, $c_{ij}$ and $c_{ijk}$ can equally well be seen as coordinates in a vector space, and the problem of finding the coefficients
 reduces to a problem of linear algebra.
 \end{enumerate}
 We have implemented the algorithm of Ref.~\cite{Gangl:2011} into a {\sc Mathematica} code, which we have applied to the function $\tilde\Phi_9(x_5,x_8,y_2,y_8,z_2,z_5)$. The result we obtain takes a strikingly simple form,
 \beq\label{eq:result}
 \Phi_9(u_1,\ldots,u_6) = {1\over\sqrt{\Delta_9}}\,\sum_{i=1}^4\sum_{g\in S_3}\,\sigma(g)\, \begin{cal}L\end{cal}_3(x_{i,g}^+,x_{i,g}^-)\,,
 \eeq
 where $\sigma(g)$ denotes the signature of the permutation ($+1$ for $\{1,c,c^2\}$, $-1$ for $\{r,rc,rc^2\}$), and where we defined
 \beq
 \begin{cal}L\end{cal}_3(x^+,x^-) := {1\over 18}\left(\ell_1(x^+) - \ell_1(x^-)\right)^3 +  L_3(x^+,x^-)\,,
 \eeq
 and
 \beq
 L_3(x^+,x^-) := \sum_{k=0}^2{(-1)^k\over (2k)!!}\,\log^k(x^+\,x^-)\,\left(\ell_{3-k}(x^+)-\ell_{3-k}(x^-)\right)\,,
 \eeq
 with
 \beq
 \ell_n(x) := {1\over2}\left(\textrm{Li}_n(x) - \textrm{Li}_n(1/x)\right)\,.
 \eeq
The arguments appearing in the polylogarithms can be written in the form $x_{i,g}^\pm := g(x_i^\pm)$, for $g\in S_3$, with
\beq\bsp\label{eq:x_plus}
&x_1^+ := \chi(1,4,7)\,,\quad x_2^+ := \chi(2,5,7)\,, \quad x_3^+ := \chi(2,4,8)\,, \quad x_4^+ := \chi(1,5,8)\,,\\
&x_1^- := \overline{\chi}(1,4,7)\,,\quad x_2^- :=  \overline{\chi}(2,5,7)\,, \quad x_3^- :=  \overline{\chi}(2,4,8)\,, \quad x_4^- :=  \overline{\chi}(1,5,8)\,,
\esp\eeq
where we defined
\beq
\chi(i,j,k) := -{\br{4\overline{7}}\br{X_iX_k}\br{X_j17}\over \br{1\overline{7}}\br{X_jX_k}\br{X_i47}}\,,
\eeq
with $\br{i\bar\jmath} = \br{i(j-1)j(j+1)}$. The function $\overline{\chi}$ is related to $\chi$ by Poincar\'e duality,
\beq
\overline\chi(i,j,k) := -{\br{\overline{4}7}\br{X_iX_k}\br{X_j\overline{1}\cap\overline{7}}\over \br{\overline{1}7}\br{X_jX_k}\br{X_i\overline{4}\cap\overline{7}}}\,.
\eeq
The function $\Phi_9$ manifestly has the cyclic symmetry. 
The reflection symmetry however needs some explanation, because $\tilde\Phi_9$ is odd under reflection. In twistor variables, $\Delta_9$ becomes a perfect square, 
and so we can remove the square root and rewrite $\sqrt{\Delta_9}$ as a rational function of twistor brackets. This procedure however introduces an ambiguity for the sign of the square root. In particular, the rational function we obtained is now odd under the reflection~\eqref{eq:twist_sym}, so that $\Phi_9$ is again even.

We stress that Eq.~\eqref{eq:result} is only valid in the region where $\Delta_9<0$. In this region, since $\chi$ and $\overline\chi$ are related by Poincar\'e duality, the function Eq.~\eqref{eq:result} is manifestly real, and we checked numerically that Eq.~\eqref{eq:result} agrees with the parametric integral representation for $\Phi_9$ given in Eq.~\eqref{H9param}. Note that, as multiple zeta values are in the kernel of the symbol map, we could a priori add to Eq.~\eqref{eq:result} terms proportional to $\zeta_2$ without altering its symbol\footnote{Note that a constant term proportional to $\zeta_3$ is excluded because of the reality condition on the function.}. The numerical agreement with the integral representation~\eqref{H9param} however shows that such terms are absent in the present case.

\section{Conclusion}
Using a differential equation method to determine the symbol of a function, and an algorithm to reconstruct the function from its symbol, we have computed analytically the one-loop non-adjacent three-mass hexagon integral in $D=6$ dimensions. Just as for the massless and one-mass hexagon integrals, the result is given in terms of classical polylogarithms of uniform transcendental weight three, which are functions of six dual conformally invariant cross-ratios. Because of the high degree of symmetry of the integral, the result is extremely compact: it can be expressed as a sum of 24 terms involving only one basic function, which is a simple linear combination of logarithms, dilogarithms, and trilogarithms.
Given the relation between one-loop hexagon integrals in $D=6$ dimensions and higher-loop amplitudes in $D=4$ dimensions, we expect that our result will help to understand the structure of $\cN=4$ SYM amplitudes and Wilson loops, particularly at two loops.

\section*{Acknowledgements}
VDD, CD and JMH are grateful to the KITP, Santa Barbara, for the hospitality while this work was carried out. CD is grateful to Herbert Gangl for valuable discussions on the symbol technique. 
This work was partly supported by the Research Executive Agency (REA) of the European Union through the Initial Training Network LHCPhenoNet under contract PITN-GA-2010-264564, by the Russian Foundation for Basic Research through grant 11-02-01196, by the National Science Foundation under Grant No. NSF PHY05-51164, and by the US Department of Energy under contract
DE--AC02--76SF00515.\\

\vspace{3mm}

{\bf Note added: }After this calculation was completed, we were informed of an independent computation of the symbols of hexagon integrals, using a different method~\cite{Spradlin}.

\appendix

\section{Special cases}
For $u_4 = u_5 = u_6 =1 $, the differential equations simplify considerably.
We have
\beq
\left[ u_{1} +  u_{1} (u_{1}-1) \partial_{1} \right] {\Phi}_{9}(u_1 , u_2 ,u_3 ,1,1,1) ={\Psi}_{8}(u_{2},u_{3},1)\,,
\eeq
where ${\Psi}_{8}(u,v,1)={\log u \log v }/(u-1)/(v-1) $, and the two cyclically related equations.
The solution is simply
\beq
{\Phi}_{9}(u_1 , u_2 ,u_3 ,1,1,1) = \prod_{i=1}^{3} \frac{ \log u_i }{u_i -1 }\,.
\eeq
The case $u_5 = u_6 = 1 $ is also very simple,
\beq
{\Phi}_{9}(u_1 , u_2 ,u_3 , u_4,1,1) =\frac{ \log u_3 }{u_3 -1 }{\Psi}_{8}(u_1 ,u_2 , u_4 ) \,.
\eeq

\section{Arguments in terms of space-time cross ratios}
In this appendix we present the expressions of the functions $x_i^+$ defined in Eq.~\eqref{eq:x_plus} in terms of the space-time cross ratios $u_i$,
\beq\bsp
x_1^+ & \,= \frac{2 u_3 \left(1-u_6\right) \left[1-u_3 u_6-u_2 \left(1-u_3 u_5 u_6\right)\right]-\left(1-u_3 u_6\right) \left(g_1-\sqrt{\Delta _9}\right)}{2 u_3 \left(1-u_6\right) \left[1-u_2-u_3 \left(1-u_2 u_5\right) u_6\right]}\,,\\
x_2^+ & \,=
\frac{2 u_1 u_3\left(1-u_6\right) \left[1-u_2 u_4-u_3 \left(1-u_2 u_4 u_5\right)\right] -\left(1-u_3\right) \left(g_6-\sqrt{\Delta _9}\right)}{2 u_1 \left(1-u_6\right)\left[1-u_2 u_4-u_3 \left(1-u_2 u_4 u_5\right)\right] }\,,\\
x_3^+ & \, = \frac{2 u_3 \left(1-u_6\right)\left[\left(1-u_2 u_5\right) \left(1-u_3 u_5\right)-u_1 \left(1-u_5\right)\right]-\left(1-u_3 u_5\right) \left(g_1-\sqrt{\Delta _9}\right) }{2 u_1 u_3 u_5 \left(1-u_6\right)\left[1-u_2 u_4-u_3 \left(1-u_2 u_4 u_5\right)\right] }\,,\\
x_4^+&\, =
-u_6\,\frac{2 u_3 \left(1-u_6\right) \left[1-u_5-u_1 \left(1-u_2 u_4 u_5\right) \left(1-u_3 u_5 u_6\right)\right]+ \left(1-u_3 u_5 u_6\right) \left(g_6-\sqrt{\Delta _9}\right)}{2 \left(1-u_6\right) \left[1-u_2-u_3 \left(1-u_2 u_5\right) u_6\right]}\,.
\esp\eeq
The variables $x_i^-$ are obtained from $x_i^+$ by replacing $\sqrt{\Delta_9}$ by $-\sqrt{\Delta_9}$.
Also, in Eq.~\ref{eq:result} we define the action of the odd permutations $g$ to include the
replacement $\sqrt{\Delta_9} \rightarrow -\sqrt{\Delta_9}$.

The twistor variables $x_i$, $y_i$ and $z_i$ rationalize the $x_i^\pm$, so that they take the form,
\beq\bsp
x_1^+ & \,= \frac{x_8}{1-y_8} \,, \\
x_2^+ & \,= -\frac{x_8(y_2-y_8)}{(1-x_8)(1-y_8)} \,,\\
x_3^+ & \,= \frac{x_8(1-y_2)}{x_5(1-y_8)} \,,\\
x_4^+ & \,= \frac{x_8 y_8}{(1-y_8)(x_5-x_8)} \,, \\
x_1^- & \,= \frac{(1-x_5)[1-x_8(1-y_2)-y_8-z_2(1-x_8-y_8)]}
 {y_2[(1-x_5)(1-y_8)-z_5(1-x_8-y_8)]} \,, \\
x_2^- & \,= -\frac{(1-x_5)(y_2-y_8)[1-x_8(1-y_2)-y_8-z_2(1-x_8-y_8)]}
  {y_2(1-x_8)[(z_2(1-x_5)-z_5)(1-y_8)+z_5 x_8 (1-y_2)]} \,, \\
x_3^- & \,= \frac{(1-y_2)(1-x_5)[(x_5(1-y_8)-x_8)(1-z_2) + x_8 y_2 (1-z_5)]}
   {y_2 x_5 [(z_2 (1-x_5)-z_5) (1-y_8) + z_5 x_8 (1-y_2)]} \,, \\
x_4^- & \,= \frac{y_8 (1-x_5) [(x_5 (1-y_8)-x_8) (1-z_2) + x_8 y_2 (1-z_5)]}
   {y_2 (x_5-x_8)[(1-x_5)(1-y_8)-z_5 (1-x_8-y_8)]} \,.
\esp\eeq
Note that these expressions correspond to a particular choice for the sign of square root.

\bibliographystyle{nb}

\end{document}